\documentclass[runningheads]{llncs}
\usepackage{graphicx}
\usepackage{xcolor}
\usepackage{amsfonts}
\usepackage{amsmath}
\usepackage{amssymb}
\usepackage{xspace}
\usepackage{wrapfig}
\usepackage{hyperref}
\usepackage[inline, shortlabels]{enumitem}

\usepackage{tikz}
\usetikzlibrary{positioning,arrows,fit}
\newcommand{\figref}[1]{Fig.~\ref{Fi:#1}}

\newcommand{\sectref}[1]{Section~\ref{Se:#1}}

\newcommand{\sat}{{\sf SAT}\xspace}
\newcommand{\sjorcidlogo}{\href{https://orcid.org/0000-0001-8070-1525}{\protect\includegraphics[scale=0.4]{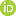}}}
\newcommand{\ecorcidlogo}{\href{https://orcid.org/0000-0002-2868-4665}{\protect\includegraphics[scale=0.4]{images/ORCIDiD_icon16x16}}}
\newcommand{\unsat}{{\sf UNSAT}\xspace}
\newcommand{\version}{{\sf 0.1}\xspace}
\newcommand{\toolname}{{\sf Pinaka}\xspace}
\newcommand{\verifox}{{\sf VerifOx}\xspace}
\newcommand{\cprover}{{\sf CProver}\xspace}
\newcommand{\symex}{{\sf Symex}\xspace}

\colorlet{sjcolor}{blue}
\colorlet{rmcolor}{green}
\begin{document}
\title{\toolname: Symbolic Execution meets \\ Incremental Solving }
\subtitle{{\small (Competition Contribution)}}
\author{Eti Chaudhary\thanks{Jury Member} \textsuperscript{\ecorcidlogo} \and Saurabh Joshi\textsuperscript{\sjorcidlogo} }
\authorrunning{E. Chaudhary \and S. Joshi}
\institute{Department of Computer Science and Engineering, \\ Indian Institute of Technology Hyderabad, India \\
\email{\{cs17mtech11029, sbjoshi\}@iith.ac.in}}
\maketitle              % typeset the header of the contribution
\begin{abstract}
Many modern-day solvers offer functionality for incremental SAT solving, which preserves the state of the solver across invocations. This is beneficial when multiple, closely related SAT queries need to be fed to the solver. \toolname is a symbolic execution engine which makes aggressive use of incremental SAT solving coupled with eager state infeasibility checks. It is built on top of the \cprover/\symex framework. \toolname supports both Breadth First Search and Depth First Search as state exploration strategies along with partial and full incremental modes. For SVCOMP 2019, \toolname is configured to use partial incremental mode with Depth First Search strategy.

\end{abstract}
\section{Verification Technique\label{Se:Approach}}

\toolname extends symbolic execution with incremental solving coupled with eager infeasibility checks.  A pure symbolic execution \cite{king1976symbolic} engine builds  
 a logical formula representing a potential execution path using symbolic values which may then be passed on to theorem-provers/solvers. An \unsat outcome from the solver implies that the verification condition will not be violated along that path, whereas a \sat outcome provides  a scenario leading to failure of an assertion during an execution along that path.
The number of paths in a program blow-up exponentially as the number of branches increases. \toolname, being a single-path symbolic execution engine, never merges two paths (i.e., diamonds).
It employs eager infeasibility checks to avoid unnecessary exploration. Rather than making queries to the solver only when a path encounters an assertion, a query is made everytime a branch is encountered to check its feasibility. Infeasible branches are not explored further. These eager checks help \toolname tremendously in reducing its search efforts.
\toolname is further powered by incremental solving \cite{hooker1993solving} offered by many state-of-the-art solvers such as MiniSAT \cite{een2006minisat}. Incremental Solving greatly benefits our technique by reducing the overhead encountered due to eager infeasibility checks. \toolname has Depth First Search (DFS) and Breadth First Search (BFS) as its search strategies. It offers two different modes of operation: Partial Incremental (PI) Mode and Full Incremental (FI) Mode.

\setlength\intextsep{0pt}
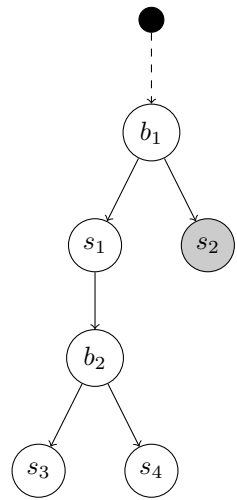
\begin{wrapfigure}{R}{0.3\textwidth}
    \centering
    \begin{tikzpicture}[nodes={draw, circle},->]
\tikzset{
 rootnode/.style={circle,draw,fill=black},
 statenode/.style={circle, draw},
 statenodeb/.style={circle,draw,fill=black!20}
 }

\node[rootnode] {}
            child{node [statenode] {$b_1$} edge from parent [->, dashed]
            child{node [statenode, solid]{$s_1$} edge from parent [->, solid]
                child{node [statenode,solid] {$b_2$} edge from parent [->, solid]
                    child{node [statenode] {$s_3$} edge from parent [->]}
                    child{node [statenode] {$s_4$} edge from parent [->]}}}
            child{node [statenodeb, solid] {$s_2$} edge from parent [->, solid]}};
                    
\end{tikzpicture}
    \caption{Branching State in a Program Graph}
    \label{Fi:incremental}
\end{wrapfigure}

In PI mode, a single solver instance is maintained along a search path. In the event that a branch is encountered, only a partial path is encoded from the current point to the previous point from which a query was made along the current path. 
For example, in \figref{incremental}, a query would be made at $b_1$. If both $s_1$ and $s_2$ are feasible, $s_2$ is put in a queue and the current solver instance is used to further explore the path starting from $s_1$. When $b_2$ is encountered, only the path from $s_1$ to $b_2$ is encoded and added in the current solver instance before making a query. If both the branches at $b_2$ are infeasible, a new solver instance is created and a path from the initial state to another symbolic state (e.g., $s_2$) in the queue is encoded and the path along that symbolic state is explored further. Essentially, a new solver instance is created every-time a backtrack happens.
Using BFS in PI mode is very memory consuming because for every symbolic state in the queue, a corresponding solver instance is retained. Running \toolname with this combination is not recommended.

In FI mode, a single solver instance is retained throughout. In \figref{incremental}, if $b_1 \rightarrow s_1$ is a feasible branch, a new activation variable $a_{b_1s_1}$ is created. Let $\phi_{b_1b_2}$ be the encoding of the path from $b_1$ to $b_2$. When $b_2$ is encountered, $a_{b_1s_1} \Rightarrow \phi_{b_1b_2}$ is added in the solver, and $a_{b_1s_1}$ is added as an assumption to enforce the path. Since the underlying SAT solvers integrated with \toolname do not allow {\em popping of a stack}, upon backtrack, $\neg a_{b_1s_1}$ is set as an assumption to disable the constraints generated by this fragment of the path.

FI mode is beneficial when the input program does not have too many paths. Otherwise, the solver becomes quite slow over time with a large 
memory footprint. For a large program with too many paths, the benefit of a lower memory footprint and speed of PI mode outweighs its overhead
of instantiating a new solver instance on every backtrack.

\begin{wrapfigure}{R}{0.3\textwidth}
\begin{verbatim}
while(x<3)
{ 
 if (y < 0 )
 { x=x+1; }
}
\end{verbatim}
\caption{Handling loops}
\label{Fi:loops}
\end{wrapfigure}

Loops are handled just like branches. Consider the program fragment given in \figref{loops}.  Assume that
along some path where $(x_1=1) \wedge (y_3=-1)$ the loop is encountered. Branch $x_1>=3$ is infeasible along this path
and will not be explored. Since $x_1<3$ is feasible, it is explored further by unrolling this iteration on-the-fly. Therefore, the
path will further add $y_3<0$. Since it is feasible, $x_2=x_1+1$ is added and feasibility of $x_2<3$ will be checked. After 
one more unrolling $x_3<3$ will be found infeasible, thus guaranteeing termination of the loop along this path. Note 
that, along a path having $y_3=2$, the loop will be non-terminating for that path. In this case, \toolname may not terminate.
Function calls, including recursion, are handled in a similar fashion by inlining a call on-the-fly.
Therefore, even though \toolname provides an option of {\tt --unwind NUM} to specify an unwinding limit, it does not mandate that a loop unwinding limit is  specified.  If a user-given unwinding limit is not sufficient to reach an assertion violation, it declares the program as {\em safe}, which 
may be unsound. To ensure soundness, we run it without any loop unwinding limit.
For {\em unsafe} programs, upon encountering the first assertion violation, \toolname terminates and reports a failure. For {\em safe} programs, however,
\toolname terminates only if all the paths of the program are terminating. 

\section{Architecture\label{Sec:Architechture}}
\toolname\ \version is built upon the \cprover\cite{cproveref}+ \symex\cite{symexref} framework\footnote{Note the specific Symex version used on which \toolname is built}. Taking a $C$ program as input, it makes use of \cprover framework APIs to convert the input C program to a GOTO program. \cprover APIs further come into play for pre-processing of GOTO-programs, witness generation, transformation passes such as setting the rounding-mode for floating-point operations, handling complex data types, etc. \toolname implements PI and FI mode and eager infeasibility checks along with BFS/DFS exploration. Apart from DFS, none of those features are present in \symex\cite{symexref}. Additionally, we make use of our forked version of the \symex repository in which we fix many bugs, especially for handling recursive procedures and ternary operators.  As of now \toolname only supports some MiniSAT-like solvers (i.e., Glucose~\cite{glucose}, MapleSAT~\cite{maplesat}) and not SMT solvers.   Once a program has been verified, a verification successful/failed outcome is generated along with the appropriate witness.
\\

\begin{figure}
    \centering
    \scalebox{0.8}{\tikzstyle{outliner} = [sensor, text width=10em, fill=red!20, 
    minimum height=6em, rounded corners, dashed]
\tikzstyle{frame}=[draw, fill=blue!20, text width=8em, 
    text centered, minimum height=2.5em, node distance=9em]
\tikzstyle{inout}=[draw, fill=green!20, text width=5em, 
    text centered, minimum height=2.5em, rounded corners, node distance=5em]
\tikzstyle{frame-special}=[draw, fill=red!20, text width=5em, 
    text centered, minimum height=2.5em, node distance=9em]
\tikzstyle{frame}=[draw, fill=blue!20, text width=8em, 
    text centered, minimum height=2.5em, node distance=7em]    
\begin{tikzpicture}
%\node (pinaka-main) [outliner]  {Pinaka-Architecture};
\node (cbmc) [frame] {\cprover/\symex Framework};
%\node (symex) [frame, right of=cbmc] {\symex Framework};
\node (pinaka) [frame-special, right of=cbmc] {\toolname};
\node (glucose) [frame, below of=pinaka, yshift=0.7em] {Solver};
\node (outline)[fit=(cbmc)(pinaka), draw, rectangle, inner sep=0.5cm, dashed,minimum height=8em]{};
\node [draw=none,fill=none,above of=outline, yshift=0.2em](stack){\textit{Framework Stack}};
\node (input) [inout, left of=cbmc, xshift=-5em]{C Program};
\node (output) [inout, right of=pinaka, xshift=5em]{Verification Outcome};
%\node (aux) [coordinate, left of=symex, xshift=-3em, yshift=-1em]{};

%\draw [<->] (cbmc) -- (symex);
\draw [<->] (cbmc) -- (pinaka);
\draw [<->] ([xshift=0.4em]pinaka.south) -- ([xshift=0.4em]glucose.north);
%\draw [<-] (pinaka) -- (aux);
%\draw [<-, dashed] (cbmc) -- (aux);
%\draw [->] (aux) -- (glucose);
\draw [->] (cbmc.south) to [out=-40, in=-150] node[above, above, xshift=0.2em, yshift=-1.5em] {GOTO Program} (pinaka.south);
\draw [->, very thick] (input) -- (cbmc);
\draw [->, very thick] (pinaka) -- (output);
\end{tikzpicture}}
    \caption{Architectural Overview of \toolname}
    \label{fig:architechture}
\end{figure}
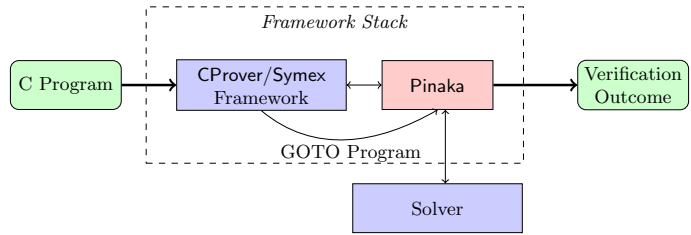

\section{Strengths and Weaknesses\label{Sec:strweak}}
Modulo the soundness of the \cprover/\symex framework and the back-end solver,  \toolname's technique is sound.
In addition, if \cprover/\symex do not have any over-approximation in modeling the $C$ constructs, then a bug reported by \toolname would indeed be a bug.

As explained in \sectref{Approach}, \toolname  can potentially be non-terminating if for some input value there is a non-terminating path. However, termination 
of verification process guarantees the termination of the underlying program (modulo the approximations introduced in modeling by \cprover/\symex) {\em if} the program is declared {\em safe} by \toolname. 
A notable strength of \toolname lies in its speed. A majority of \toolname's verification outcomes were obtained under a mere limit of 10 seconds. A clear display of the same can be seen in the ReachSafety-Floats category, where \toolname came in second with $1500$ seconds CPU time \cite{svcomp19}, as compared to other tools that share a similar score but require 4 to 10 times more CPU time.

One major weakness of the current version \toolname is a lack of techniques for loop invariants. Even with eager infeasibility checks and incremental solving, there is still a need for more loop-directed abstraction based approaches. Furthermore, support for handling multi-dimensional arrays is still lacking.

\section{Tool Setup and Configuration\label{Sec:setup}}
\toolname\ \version is available for download at \url{https://github.com/sbjoshi/Pinaka}. The repository contains a description of \toolname's working along with all the necessary configuration files required to run \toolname SVCOMP style. All the instructions are listed in a stepwise manner in the README.md file. Although \toolname is built on top of the \cprover/\symex framework, the binary itself is sufficient and the tool does not require any additional pre-requisites. \toolname has been tested on Ubuntu 18.04.

\toolname\ \version submitted for SVCOMP 2019 runs DFS in PI Mode as for SVCOMP benchmarks we found this combination the best. No loop unwinding limit was specified to retain soundness.
For SVCOMP'19~\cite{SVCOMP19Report,svcomp19} it uses Glucose-Syrup (Glucose-4.1) \cite{glucose} as its solver back-end.
Tool's default search strategy, i.e., DFS may be overridden by providing \texttt{--bfs} option. Similarly, a default of FI mode may also be overridden by providing \texttt{--partial-incremental} option. Other additional options may be explored from the \texttt{--help} menu.
The set of global parameters passed to the tool are:
\begin{enumerate*} [(1)]
\item \texttt{--graphml-witness}: to specify the witness file to be generated, 
\item \texttt{--propertyfile}: to specify the property file, 
\item \texttt{--32/--64}: to define the architecture to be used.
\end{enumerate*}
\toolname\ \version participated in all ReachSafety subcategories \textit{except} ReachSafety-Sequentialized, and also participated in NoOverflows and Termination meta-categories in SVCOMP 2019. 

\section{Software Project and Contributors\label{Sec:contri}}
\toolname is  a result of very heavy code rewriting and refactoring of \verifox~\cite{mukherjee2016equivalence} (developed by Saurabh Joshi) with a lot of feature additions and bug fixes.
\toolname is developed at Indian Institute of Technology, Hyderabad, India. 
It is available at \url{https://github.com/sbjoshi/Pinaka} under BSD License.
The authors acknowledge the financial support from DST, India under SERB ECR 2017 grant.
%Contact info (web page of the project, people involved in the project)
% Information about the software project, licensing, development model, institution that hosts the software project, acknowledgement of contributors

\bibliographystyle{splncs04}
\bibliography{refs}

\begin{thebibliography}{10}
\providecommand{\url}[1]{\texttt{#1}}
\providecommand{\urlprefix}{URL }
\providecommand{\doi}[1]{https://doi.org/#1}

\bibitem{SVCOMP19Report}
Beyer, D.: Automatic verification of c and java programs: Sv-comp 2019. In:
  Proc.\ TACAS, part 3. LNCS~11429, Springer (2019)

\bibitem{cproveref}
Cprover homepage. \url{http://www.cprover.org}, last accessed 10 Feb 2019

\bibitem{een2006minisat}
Een, N., S{\"o}rensson, N.: Minisat v2. 0 (beta). Solver description, SAT race
  \textbf{2006} (2006)

\bibitem{glucose}
Glucose's homepage. \url{http://www.labri.fr/perso/lsimon/glucose/}, last
  accessed 10 Feb 2019

\bibitem{hooker1993solving}
Hooker, J.N.: Solving the incremental satisfiability problem. The Journal of
  Logic Programming  \textbf{15}(1-2),  177--186 (1993)

\bibitem{king1976symbolic}
King, J.C.: Symbolic execution and program testing. Communications of the ACM
  \textbf{19}(7),  385--394 (1976)

\bibitem{maplesat}
Maplesat homepage. \url{https://sites.google.com/a/gsd.uwaterloo.ca/maplesat/},
  last accessed 10 Feb 2019

\bibitem{mukherjee2016equivalence}
Mukherjee, R., Joshi, S., Griesmayer, A., Kroening, D., Melham, T.: Equivalence
  checking of a floating-point unit against a high-level c model. In: FM (2016)

\bibitem{svcomp19}
\uppercase{SVCOMP} 2019 results.
  \url{https://sv-comp.sosy-lab.org/2019/results/results-verified/}, last
  accessed 4 Jan 2019

\bibitem{symexref}
Symex repository.
  \url{https://github.com/diffblue/symex/tree/9b5a72cf992d29a905441f9dfa6802379546e1b7},
  last accessed 10 Feb 2019

\end{thebibliography}
\end{document}